\documentclass[twoside,a4paper,11pt]{sca}
\usepackage[pdftex]{graphicx}
\usepackage{hyperref}
\usepackage{movie15}
\begin{document}
\pagenumbering{arabic}
\pagestyle{myheadings}
\thispagestyle{empty}
{\flushright\includegraphics[width=\textwidth,bb=90 650 520 700]{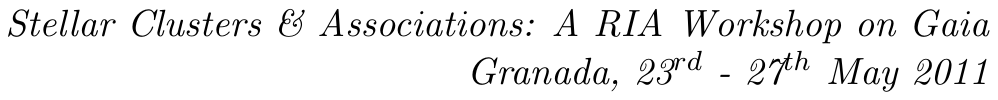}}
\vspace*{0.2cm}
\begin{flushleft}
{\bf {\LARGE
%
Improved distances to several Galactic OB associations
%
}\\
\vspace*{1cm}
%
N. Kaltcheva$^{1}$
and
V. Golev$^{2}$
%
}\\
\vspace*{0.5cm}
%
$^{1}$
Department of Physics and Astronomy, University of Wisconsin Oshkosh,
  800 Algoma Blvd., Oshkosh, WI 54901, USA (kaltchev@uwosh.edu)\\
$^{2}$
Department of Astronomy, Faculty of Physics, St Kliment Ohridski University of
Sofia, 5 James Bourchier Blvd., BG-1164 Sofia, Bulgaria (valgol@phys.uni-sofia.bg)\\
%
\end{flushleft}
%
\markboth{
Galactic OB associations
}{ 
%
Kaltcheva \& Golev  
%
}
\thispagestyle{empty}
\vspace*{0.4cm}
\begin{minipage}[l]{0.09\textwidth}
\ 
\end{minipage}
\begin{minipage}[r]{0.9\textwidth}
\vspace{1cm}
\section*{Abstract}{\small
%

Based on $uvby\beta$ photometry we study the structure of several Galactic
star-forming fields.  Lac OB1 is a compact association at 520$\pm20$ pc
spatially correlated with a region of intense H{\sc ii} emission in
Sh2-126. Lod\'{e}n 112 is a compact OB group at 1630$\pm82$ pc, probably connected to
an extended feature of OB stars located toward the Carina tangent. 
The field toward Car OB1 is complex and 
likely contains apparent concentrations representing parts of long segments
of the Carina arm projected along the line of sight. Within the classical Mon
OB2 association we separate a relatively compact group
at 1.26 kpc, that is spatially correlated to the Monoceros Loop SN remnant.
%

\normalsize}
\end{minipage}
%
%
%
\section{Introduction \label{intro}}
The Galactic OB-associations offer a unique opportunity to study the
influence of massive stars on the interstellar matter. 
A reconstruction of the star-formation history of many Galactic fields should be
possible once the spatial distribution of the young stars is reliably
determined. Despite the extensive efforts to improve and unify the distances
to the young stellar groups in the Milky Way (MW), discrepancies still remain
in the published studies for a large number of fields. Many of the present
distance estimates are based to a large extent on preliminary distance
calibrations, broad-band photometry, or absolute magnitudes ($M_V$) obtained
via spectral and luminosity type (MK classification). On the other hand, the
$uvby\beta$ photometric system provides  $M_V$ and colour
excess determinations for early-type stars in excellent agreement
with the $Hipparcos$ parallaxes \cite{kaltcheva98}.
\begin{figure}
\center
\includegraphics[scale=0.27]{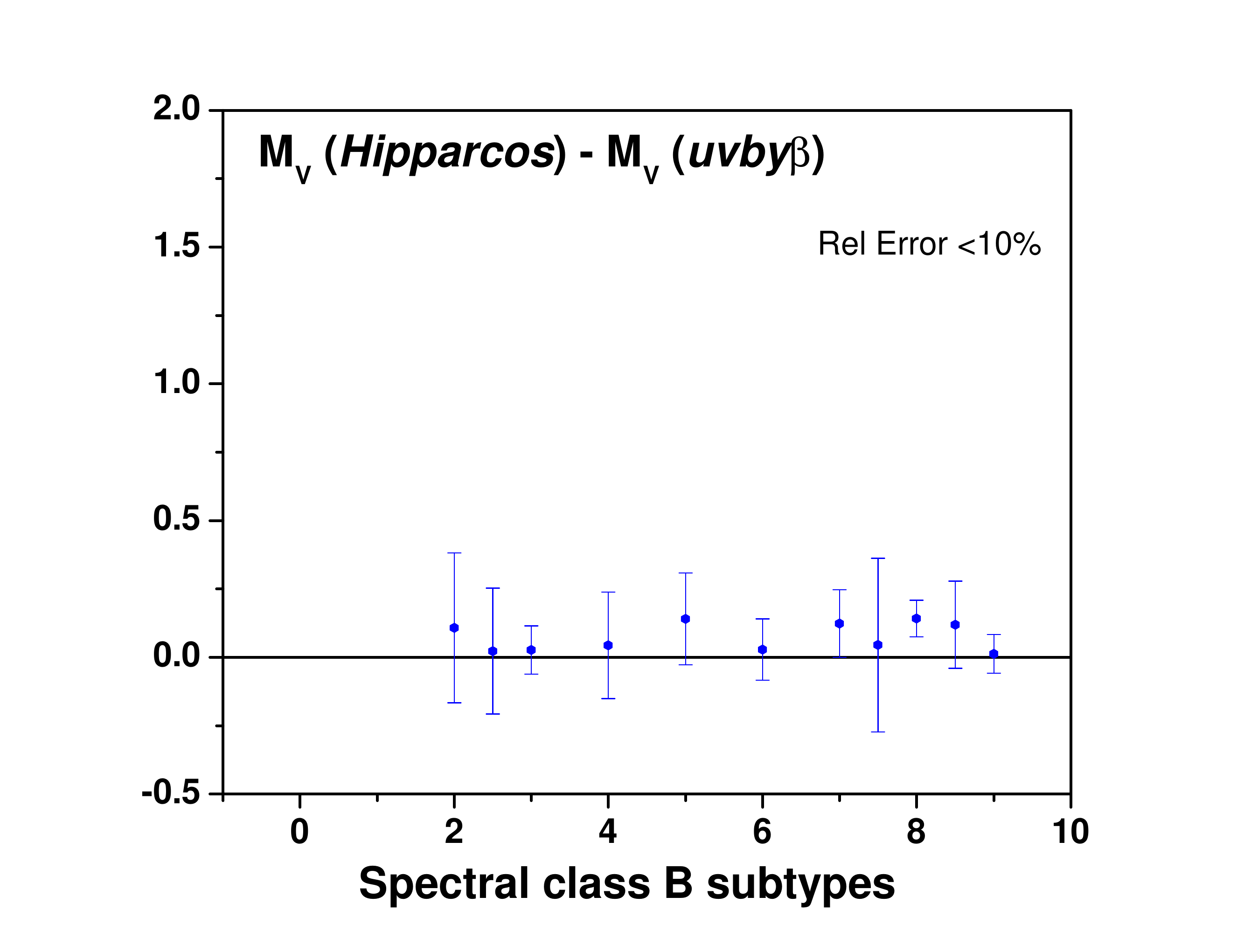} ~
\includegraphics[scale=0.27]{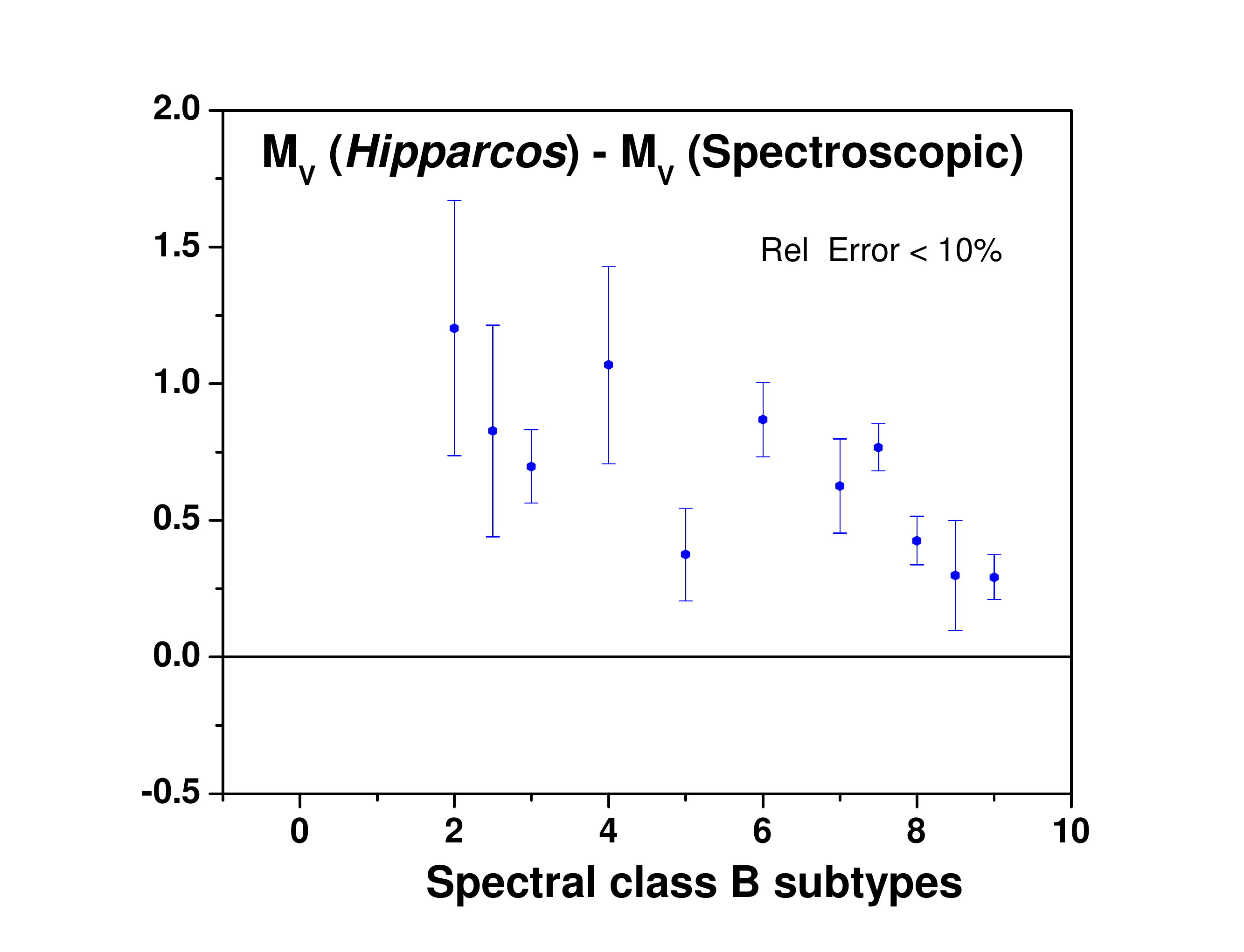} 
\caption{\label{fig1} Comparisons of  $uvby\beta$ $M_V$ (left) and
MK-based $M_V$ (right) to the $Hipparcos$-based $M_V$.}
\end{figure}

Figure~\ref{fig1} represents the comparison of $uvby\beta$ $M_V$ and MK-based
$M_V$ to the $Hipparcos$-based $M_V$, pointing out to a possible
over-estimation of stellar distance when relying on a MK-based determination.
Since the distances to the Galactic OB associations are based mainly on
individual stellar distances and rarely on main-sequence fitting, applying
$uvby\beta$ photometry should lead to a significant improvement for many OB
groups in the MW.

Our deriving of $uvby\beta$ photometric distances utilizes the
intrinsic colour calibrations of Crawford \cite{crawford78} and Kilkenny \& Whittet \cite{kilkenny85} and
the luminosity calibration of Balona \& Shobbrook \cite{balona84} and takes into account possible
stellar emission and mis-classification \cite{kaltcheva00}. The expected
errors for one star are about 12 \% for luminosity classes III-IV and about
18-20~\% for luminosity classes I and II. Photometric $uvby\beta$ distances
derived in this way provide the same impression of the star-forming field's
structure as the improved $Hipparcos$ parallaxes (cf. \cite{kaltcheva07}). 

In this contribution we present improved distance estimates for four Galactic
OB associations.

\section{Lac OB1}

Lac OB1 (Lac~OB1b, \cite{blaauw58}) is a nearby notable clustering of
early-type stars near 10 Lacertae that initially gained attention because of
the expanding motion of its members. Based on the derived $uvby\beta$
photometric distances used in conjunction with the photometric diagrams we
identify Lac OB1 as a compact group of 12 low-reddened main-sequence stars
located at a distance 520$\pm20$ pc in the direction $l=96.4^\circ, b=-16.6^\circ$ (see \cite{kaltcheva09} for details). The available radial velocity and
proper motion measurements support the impression that this is a real
group. For these 12 stars, the recalculated $Hipparcos$ parallaxes
\cite{vanleeuwen07} are in excellent agreement with the photometric
$uvby\beta$ parallaxes. The photometric distance of the O9V star 10~Lac (HD
214680) is estimated to be 715$^{+107}_{-92}$ pc. Although this estimate is
considerably larger than the one based on $Hipparcos$, the agreement for this
stars is better with the recomputed $Hipparcos$ parallax \cite{vanleeuwen07}
which yields a distance of $529^{+70}_{-50}$ pc in comparison to the original
$Hipparcos$ estimate of $325^{+82}_{-55}$ pc.

\begin{figure}
\center
\includegraphics[scale=0.5]{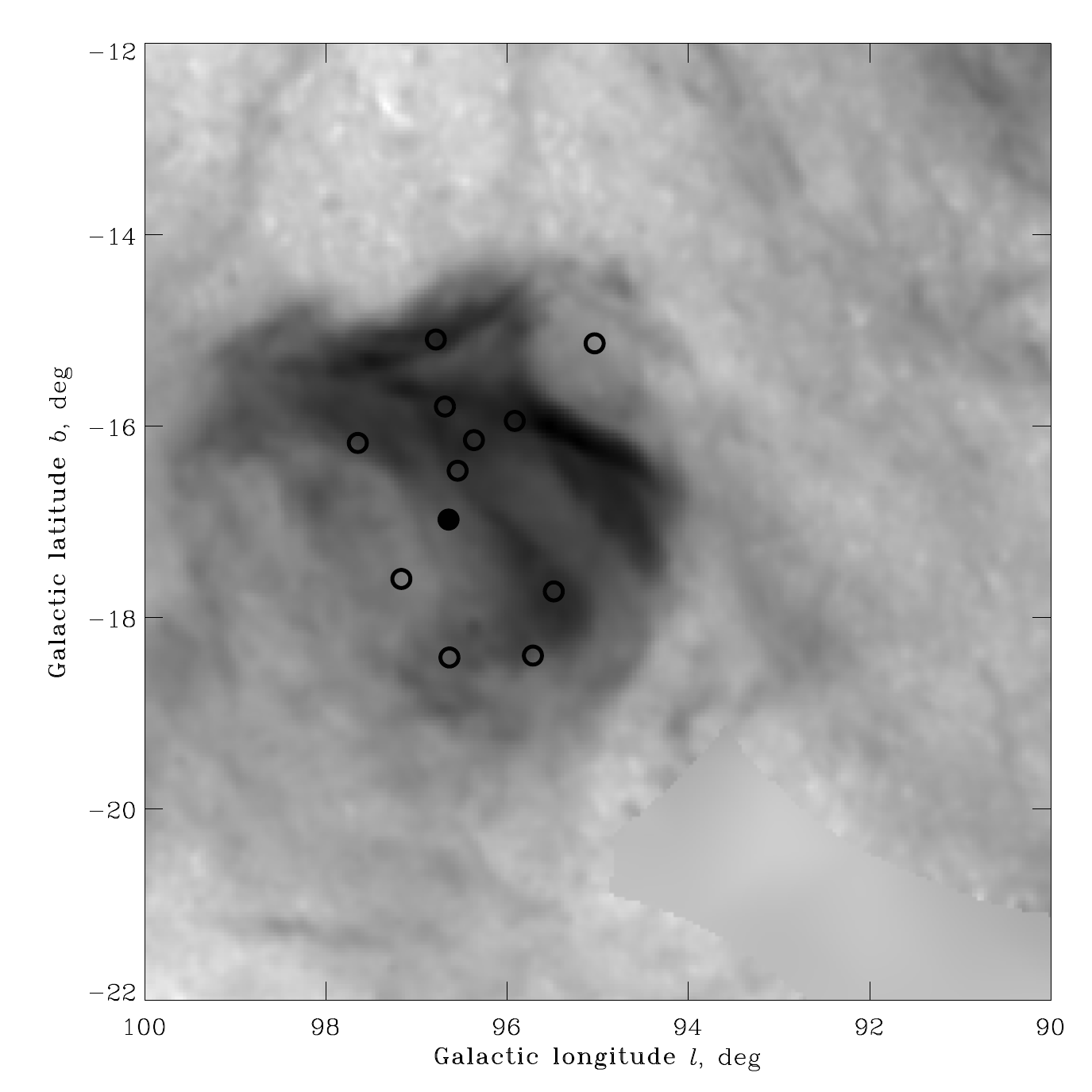} ~
\includegraphics[scale=0.5]{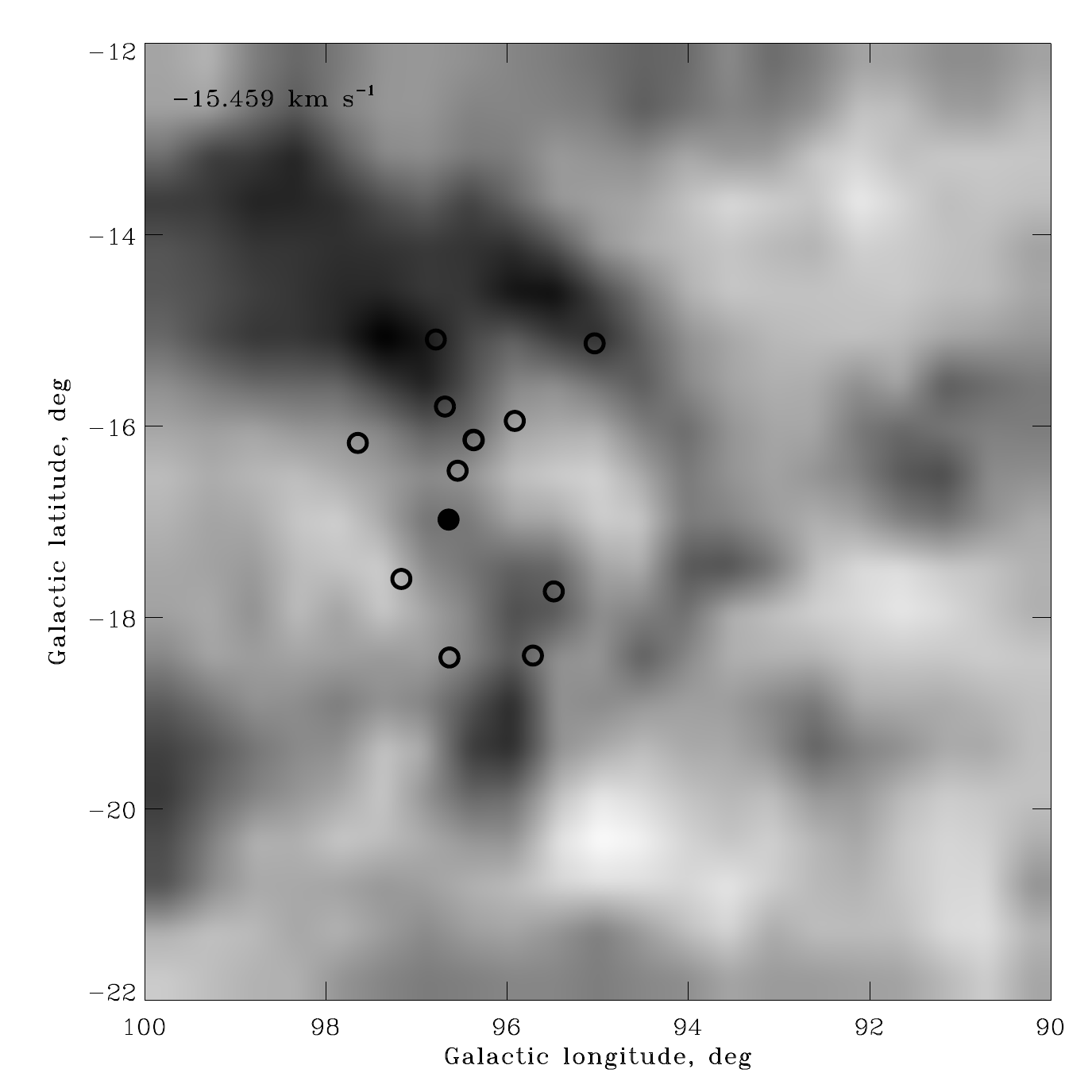} 
\caption{\label{fig2} The stars of Lac OB1 overploted on the distribution of
  the H{\sc ii} emission in Sh2-126 (left) and H{\sc i} (right). 10 Lac is shown with filled symbol.}
\end{figure}

Figure~\ref{fig2} presents the distribution of H{\sc ii} intensity in units of
Rayleighs and brightness temperature distribution of H{\sc i} at velocity
channel $-15.5$ km s$^{-1}$ toward Lac OB1, with Lac OB1 stars
superimposed. Here and on  all further  figures the H{\sc ii} data are taken from \cite{finkbeiner03} via the $SkyView$ interface \cite{McGlynn98}, and the H{\sc i} data are taken from Leiden/Argentine/Bonn (LAB) Survey of Galactic HI \cite{kalberla05}.
A correlation of the stars' location with the regions of
intense H{\sc ii} emission in Sh2-126 \cite{sharpless59} is noticeable. On the
other side, the distribution of neutral hydrogen shows a deficiency (most obvious in the selected velocity channel), also
correlating with the location of the stars (see also \cite{cappadenicolau90}).

\section{The field of Lod\'{e}n 112}

Lod\'{e}n 112 is identified as a poor, but compact cluster candidate. Based on $uvby\beta$ photometry we obtained true DM = 11.06$\pm$0.12(s.e.) and average color excess $E(b-y)$ = 0.5$\pm$0.03(s.e.) \cite{kaltcheva11}. This corresponds to a distance of 1630$\pm$82 pc, which is
significantly smaller than the presently adopted 2500 pc (WEBDA). In our
$uvby\beta$ sample there are several other early B stars located at that exact
distance. The photometric distances and available proper motions allowed us to
identify a group of about 10 early B stars that could represent a new OB
association at coordinates $282^\circ\!<l<285^\circ\!$,
$-2^\circ\!<b<2^\circ\!$ which appears to be connected to Lod\'{e}n 112
cluster candidate (see a poster by Kaltcheva \& Golev at this meeting for more details).

\section{Car OB1}
\begin{figure}
\center
\includegraphics[scale=0.5]{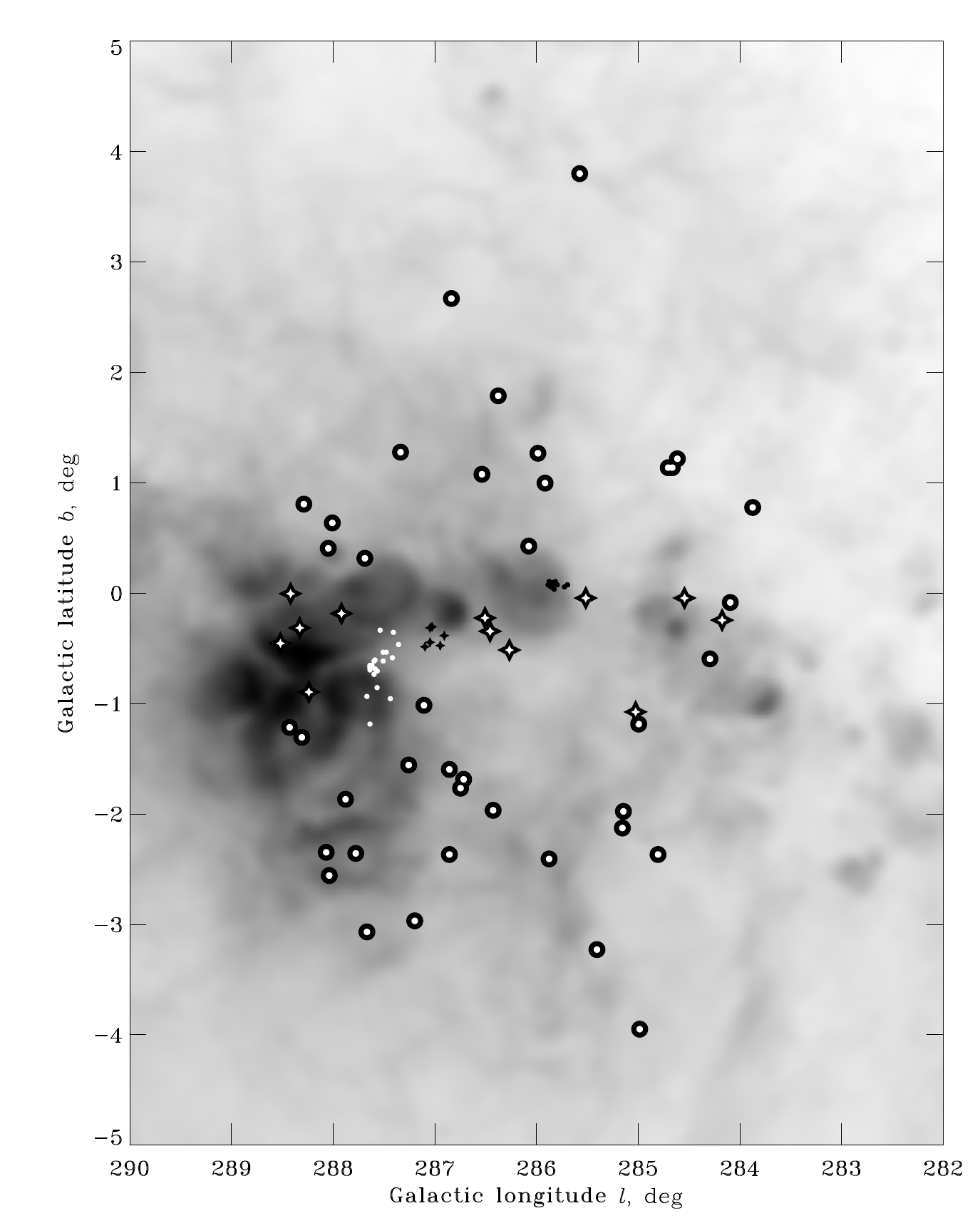} ~
\includegraphics[scale=0.5]{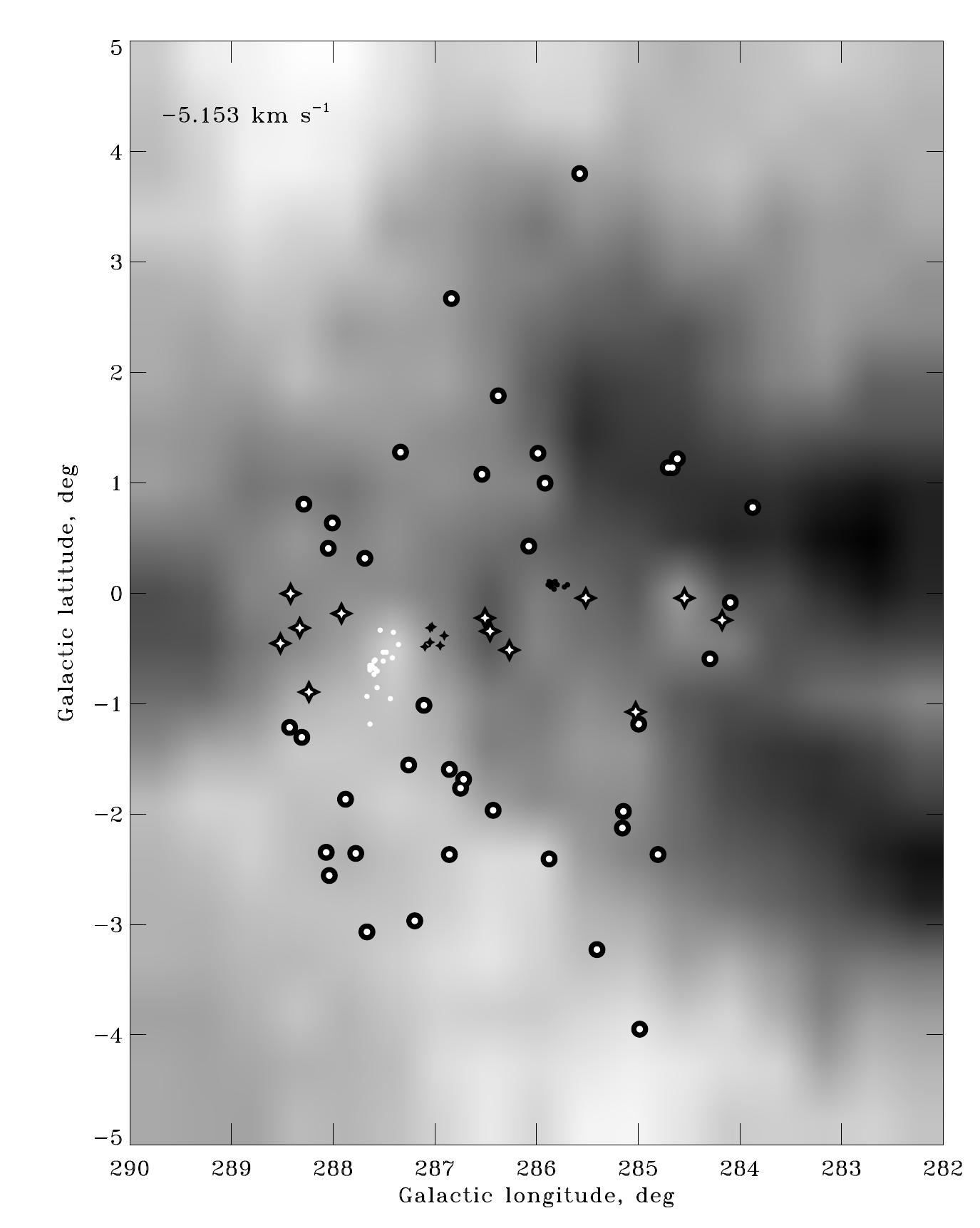} 
\caption{\label{fig3} The stars in the Car OB1 field overplotted on the
  distribution of the H{\sc ii} emission  (left) and the H{\sc i} velocity chanel image at $v=-5$ km s$^{-1}$  (right).}
\end{figure}

According to \cite{humphreys78} the Car~OB1 association is located at 2.5 kpc
toward $284^\circ\!<l<288^\circ\!$, $-2.2^\circ\!<b<0.9^\circ\!$ and has
an average radial velocity of $-5$ km s$^{-1}$ (see \cite{melnik09}). In their revision
of the list of Galactic OB associations Melnik \& Efremov \cite{melnik95} break down Car~OB1
into five groups at distances between 2.2 and 2.8 kpc.  Figure~\ref{fig3}
presents the H{\sc i} velocity chanel image at $v=-5$ km s$^{-1}$ and the
distribution of H{\sc ii} toward Car OB1. Only stars
intrinsically brighter than $-$3 mag with $uvby\beta$ distances available are shown.
Different symbols are used to denote the apparent groups that could be
separated based on our sample. The group marked with  white dots (23 stars)
is located at  average coordinates $l=287.56^\circ\!$, $b=-0.67^\circ\!$
in  direction of the Car~1E group from the list of \cite{melnik95}, in the
vicinity of $\eta$~Car. According to the distances
obtained here, these stars are spread out between 2204 and 6404 pc, with an
average of 3728$\pm956$ pc. Another apparent concentration (small black
plus-symbols, six stars) is found at $l=287.01^\circ\!$, $b=-0.41^\circ\!$ and an
average distance of 3308$\pm2090$ pc. The significant spread in distance
suggests that these are not physical groups. The third apparent grouping,
marked with black dots (16 stars), is found at $l=285.83^\circ\!$,
$b=0.071^\circ\!$ (this is Car~1B of \cite{melnik95}). These stars are well
grouped at 2583$\pm70$(s.e.) pc (see \cite{kaltcheva10} for more details). It
seems that these apparent groups are located along the edge of the Carina arm.
The apparent field stars along the edge of the arm are marked with large star-symbols
whereas the other field stars are marked with large circles.

This region has a complex structure and does not harbour just one stellar
association. It is highly likely that some of the apparent concentrations
represent parts of long segments of the Carina arm projected along the line of
sight near the Carina tangent.

\begin{figure}
\center
\includegraphics[scale=0.5]{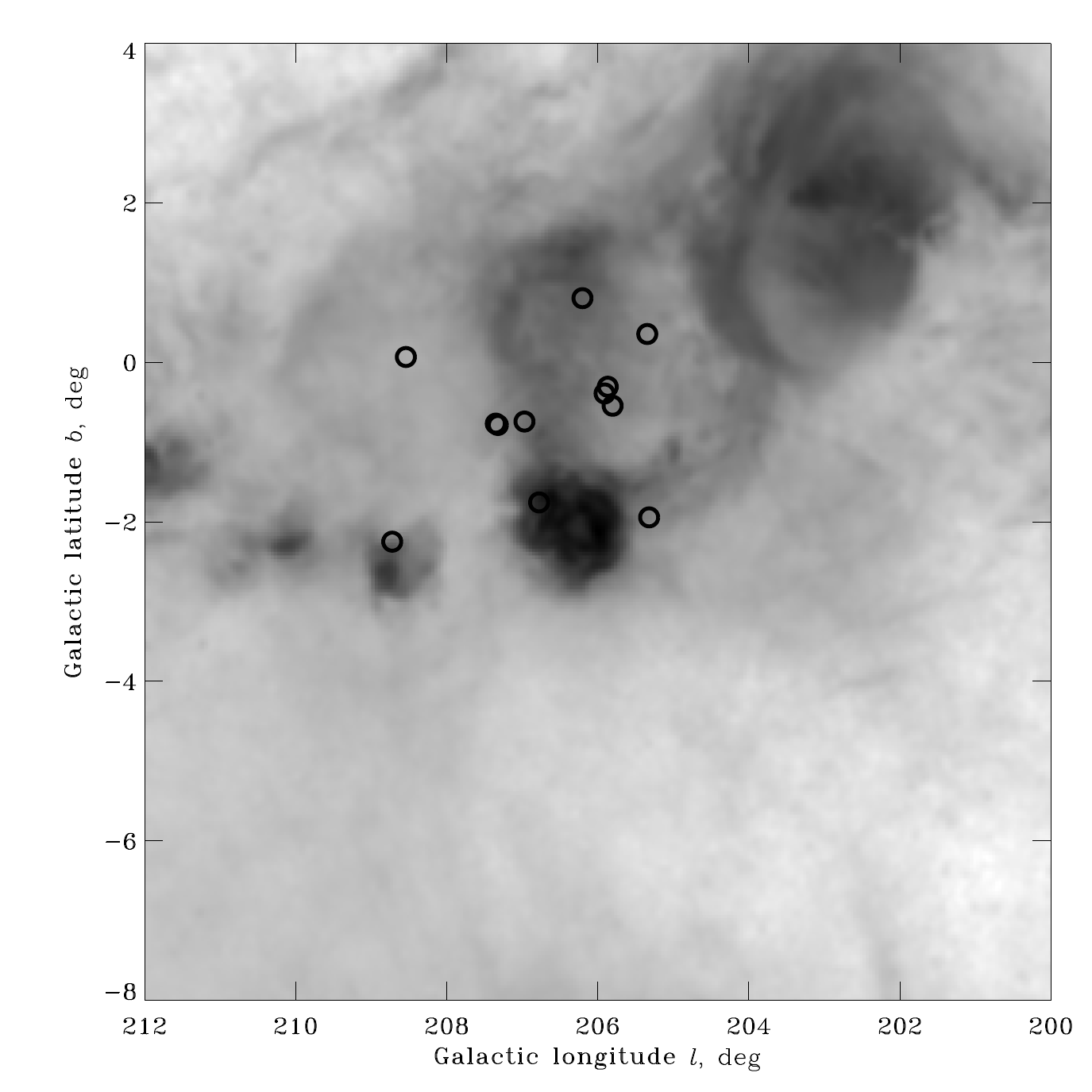} ~
\includegraphics[scale=0.5]{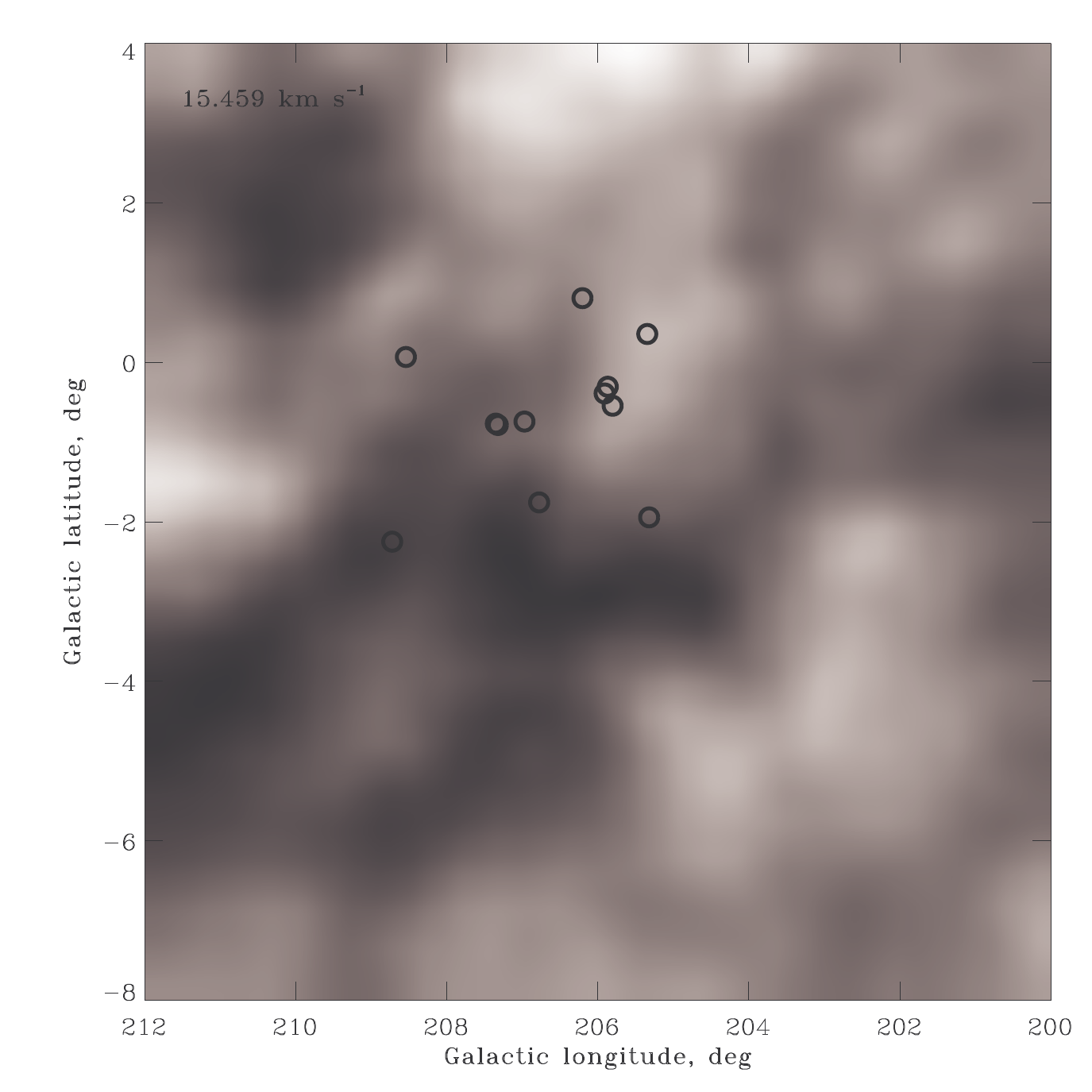} 
\caption{\label{fig4} The field of Mon OB2. The stars of the group at 1.26 kpc
  are overploted on the distribution of the H{\sc ii} emission (left)  and
  the H{\sc i} velocity chanel image at $v=15.5$ km s$^{-1}$  (right).}
\end{figure}

\section{Mon OB2}
In light of the growing importance of Northern Monoceros in the study of
star-formation, precise distance estimates to the apparent structures of young
stars in the field are most important for a variety of reasons. Northern
Monoceros is dominated by the Rosette Nebula, thought to be spatially
correlated to the young open cluster NGC~2244 and the large Mon OB2
association \cite{ruprecht66}. Precise $uvby\beta$ stellar distances and
reddening are derived for a sample of more than 200 O-B9 stars in a $12^\circ\,\times\,12^\circ$ field centered on NGC~2244 \cite{kaltcheva10a}.  According to
this sample, the classical Mon OB2 association, previously thought to be at
1.6 kpc, is represented by a relatively compact group at 1.26 kpc in the
vicinity of NGC~2244 and a layer of massive stars between 1.5 and 3 kpc spread
across the entire field.

Figure~\ref{fig4} presents the distribution of H{\sc ii} toward the group at
1.26 kpc and
the H{\sc i}  velocity chanel image at 
$15.5$ km s$^{-1}$ (at which the lack of H{\sc i} in the vicinity of this  group is more obvious). 
New estimations of the temperature, brightness, spectral index, and
distance to the Monoceros Loop SN remnant were recently reported by Borka Jovanovi{\'c} \& Uro{\v s}evi{\'c}
\cite{borkajovanovic09}. They stressed the influence of molecular cloud
on Monoceros SN remnant and calculated a distance of $1250 \pm 190$ pc to the Loop. 
This new distance estimate is in excellent agreement with the distance to the group at 1.26 kpc.

\section{Concluding Remarks}
As already mentioned, a better understanding of the phenomena associated with
the interaction between young massive stars and their surrounding ISM in the
fields of the Galactic OB associations should be possible once their structure
is reliably established. A multi-wavelength approach analyzing the
distribution of the massive OB stars, ionized and neutral material, and that
of the interstellar dust would allow us to better understand the different
components of the ISM and the
interactions among them.

%
\small  
%
\section*{Acknowledgments}   
%
This work is supported by the National Science Foundation grant
AST-0708950.  N.K. acknowledges support from the SNC Endowed Professorship
at the University of Wisconsin Oshkosh.  V.G. acknowledges  support by
the Bulgarian National Science Research Fund grants DO 02-85/2008 and
DO 02-362/2008. This research has made use of the SIMBAD database, operated
at CDS, Strasbourg, France. We acknowledge the use of NASA's {\em SkyView}
facility (http://skyview.gsfc.nasa.gov) located at NASA Goddard Space Flight
Center  \cite{McGlynn98}. 
%

%
\end{document}